\documentclass[12pt]{article}

\usepackage{graphicx}
\usepackage{amssymb}
\usepackage{verbatim}
\begin{document}

\title{Quantum mechanics, objective reality, and the problem of consciousness}

\author{Ranjan Mukhopadhyay \\
Department of Physics, 
Clark University, 
Worcester, MA 01610. \\
Email: ranjan@clarku.edu }

\maketitle

\begin{abstract}
      The hard problem in consciousness is the problem of understanding how physical processes in the brain could give rise to subjective conscious experience. In this paper, I suggest that in order to understand the relationship between consciousness and the physical world, we need to probe deeply into the nature of physical reality. This leads us to quantum physics and to a second explanatory gap: that between quantum and classical reality. 
 I will seek a philosophical framework that can address these two gaps simultaneously. Our analysis of quantum mechanics will naturally lead us to the notion of a hidden reality and to the postulate that consciousness is an integral component of this reality. The framework proposed in the paper provides the philosophical underpinnings for a theory of consciousness while satisfactorily resolving the interpretation problem in quantum mechanics without the need to alter its mathematical structure. I also discuss some implications for a scientific theory of consciousness. 
\end{abstract}

\section{Introduction}
At the heart of our scientific endeavor to understand the world, exist two explanatory gaps relating to the nature of reality and of consciousness. The first, better known, gap concerns the relation between subjective conscious experience and physical neurological processes in the brain. The relationship between the quantum micro-world and the classical world that we experience constitutes the second gap. As I will discuss, these explanatory gaps appear to be fundamental and pose a barrier to our understanding of the nature of the world and our place in it. The thesis that I seek to develop in this article is that these gaps are inter-related and cannot be addressed in isolation. The purpose of this paper is to examine critically these explanatory gaps and to discuss the possible nature of a self-consistent philosophical framework that could simultaneously address them. 

    The first gap is well-known, rooted as it is in the mind-body problem with a long history in philosophy and metaphysics, both in the East and the West. 
It is intimately related to the puzzle of how consciousness fits in with our notion of the physical universe; it is my attempt to resolve this puzzle that constitutes the central motivation for the paper. The second gap, while relatively modern, is however no less fundamental.  It lies at the heart of our interpretation of quantum mechanics, and challenges our very notion of an objective observer-independent external reality. To quote Arthur Fine (1996) ``Realism is dead. ... Its death was hastened by the debates over the interpretation of quantum theory, where Bohr's nonrealist philosophy was seen to win out over Einstein's passionate realism. Its death was certified, finally, as the last two generations of physical scientists turned their backs on realism and have managed, nevertheless, to do science successfully without it.''

     The notion of realism, however, lies at the heart of human scientific enterprise, and scientists in most fields of science take it for granted that they want to discover some aspect of how the world really works. The notion of an objective external observer-independent reality is also central to our day-to-day existence and activity. Thus I propose a commitment to realism, implying that I will assume, at the outset, the existence of a physical reality that is referenced by our physical theories, including quantum mechanics, and that this reality is objective in the  sense of existing independent of us and our perceptions. I will also use the phrase `empirical reality' somewhat loosely to refer to the physical world that we experience without assuming {\it a priori} its objectivity; indeed we would like the precise relation between `empirical reality' and objective physical reality to emerge from our analysis. While my assumption of realism might run counter to some of the accepted wisdom in quantum mechanics, I will argue later that the true issue with quantum mechanics is not about realism versus anti-realism, rather it concerns the nature and role of consciousness.   

          Since so much has been written on the possible connection between quantum mechanics and consciousness, let me clarify my position at the outset. I will do so in two parts. We will find that in order to make sense of quantum mechanics we need to address both the question of what quantum mechanics is telling us about the nature of reality, and, also, how to relate `quantum reality' to our experience of external reality. The second question is crucial, the entire measurement problem in quantum mechanics revolves around it, and it is in seeking to address this question that we encounter consciousness. But why does the interpretation of quantum mechanics, in turn, have any bearing on our understanding of consciousness? I am  not proposing a quantum theory of consciousness. On the contrary, I will argue that quantum mechanics itself points us away from physicalism: the view that all that exists is physical, that is, describable by physics and physical laws. I will let quantum mechanics guide us towards an alternative self-consistent philosophic framework, where consciousness should be treated as being just as fundamental as physical reality (my position will, however, be distinct from Cartesian dualism).

\section{The hard problem of consciousness}
    
 I begin with an examination of the first gap concerning the relationship between consciousness and the brain. The ``common sense'' view today among many scientists, is that mental processes and consciousness correspond, in some sense, to physical activities in the brain. Maybe consciousness is an emergent property of complex highly organized matter in the form of the brain. Such an approach appears sensible; after all, there clearly is sufficient evidence that chemical changes in the brain affect our mental states. Moreover, neurological studies provide direct evidence that different mental and emotional states can be distinguished in terms of the patterns of neural activity (neuronal firings) in the brain (Laureys et al., 2008). Yet there appears to exist a gap between subjective experience and physical neurological processes in the brain as highlighted by philosophers such as Nagel, Jackson, and Chalmers (Nagel, 1974; Jackson, 1982; Chalmers 1995, 1996, 2002b). David Chalmers has distinguished between two types of problems related to consciousness: the easy problems and the hard problem. The hard problem is the problem of how and why we have qualia (the subjective aspects of our experience or sensations). How do objective physical processes in the brain give rise to subjective sensations that are only accessible in the first person? Chalmers contrasts this with the so-called easy problems, which though maybe not easy to solve in practice, can in principle be solved by understanding the physical mechanisms that can perform that function; examples include the understanding of the ability to discriminate, categorize, and react to environmental stimuli and the integration of information by a cognitive system.

Much has been written about the nature of the hard problem, though specialists differ widely on the character and implications of the gap (see Chalmers, 2002a, for a variety of positions). Of particular relevance are the books by Chalmers (Chalmers 1996, 2010) that discusses the gap and builds a strong case against a physicalist interpretation of consciousness. Rather than reproducing these arguments, let me simply highlight here one aspect of the gap by focusing on the sensation of color. We have a certain subjective sensation of `blue' when we see the color blue, and a different sensation of `green' when we see the color green. What dictates the quality of the sensations? I could study all the physical processes involved, from the activation of appropriate photoreceptors in the eye to the pattern of firing of neurons in the brain, but where in all this lies the actual personal experience of the sensation of color? As argued by Nagel in his article ``What is it Like to Be a Bat?'' (Nagel, 1974), consciousness has essential to it a subjective character, a what it is like aspect. And it is not at all clear that this subjective first personal experience can be entirely represented within an objective physical framework. I will occasionally use the phrase `phenomenal consciousness' to highlight this subjective aspect of consciousness.

   Since challenging a physicalist interpretation of phenomenal consciousness has sometimes been compared to the vitalist position held by some biologists in the early twentieth century, it is important to highlight the essential difference between the two positions. While Chalmers has successfully done so in his book (Chalmers, 1996), due to the central importance of these arguments for this paper, I will briefly highlight the difference in terms of a simple, though maybe somewhat technical, thought experiment. Imagine developing an extremely complex and powerful computer simulation of a living cell; in the simulation we would input all the physical constituents (all the atoms, for example), all the physical interactions between the constituents, all the relevant structural information, and then let the simulation run. If the simulation can demonstrate all the physical functions that a cell performs and that we associate with life, then there is no further reason for vitalism. There are very good reasons why developing such a simulation is not feasible and probably not a good idea, but here I am purely making a logical point and will not worry about the practicalities of such a simulation. Now imagine developing, similarly, a simulation of the entire brain (at whatever level of detail that is appropriate, also if we need to include all our sense organs in the simulation, so be it) in order to understand how conscious experience emerges from the brain and its physical activities. If such a simulation were successful, it should be able to predict the relation between the quality of subjective experience and physical processes in the brain. It should be able to predict how, if we change some structural aspects of the brain, our subjective experiences get modified, as appears to be the case with synaesthesia (Ramachandran et al., 2003). Again not worrying about the practicalities of such a simulation, the problem that we will encounter is the following: even if the simulation could successfully capture the causal relationship between all physical processes in the brain, how would we represent qualia in our physical simulation? We encountered no problem of a similar nature with the simulation of a living cell. A little thought should convince us that for the possible success of the simulation in explaining consciousness, we are left with the options of either denying the phenomenological reality of conscious experience altogether, or we could develop an elaborate set of rules relating physical processes to conscious experience. The second position, it can be argued, is not a physicalist position since these rules would then have to be treated as fundamental and will not themselves follow from physical laws (we could probably characterize this as a form of property dualism). Thus the only viable position compatible with physicalism appears to be the denial of the phenomenal reality of qualia and thus the need to explain subjective experience, a position I find unsatisfactory considering that subjective sensations and perceptions are what we experience directly and we experience the external world only indirectly through our subjective sensations/perceptions. In this article I will  assume the existence of phenomenal experience and consequently rule out this possibility. For a physicalist, there could be further options that are available, such as the possibility that some new hitherto unimagined laws of physics will negate the above argument, and while it is difficult to rigorously rule out all such possibilities, they appear mysterious to me at present.  

    An alternative to physicalism, that of radical idealism, as proposed in the eighteenth century by Bishop Berkeley, which posits that consciousness is primary and matter (or the external world) is secondary, a creation of the mind, also appears problematic in explaining inter-subjective agreement between observers; it is, moreover difficult to sustain as a scientist. If the external world is a construct of the mind, why do we need to design and execute carefully controlled experiments to study nature, as is the norm in science?  A historically popular third alternative, characterized in modern language as substance dualism, and closely associated with the philosophy of Ren{\'e} Descartes, holds that the mind is a nonphysical substance in a physical body. This alternative seems rather mysterious as well as it is not clear how the supramaterial consciousness acts upon matter. If the mind and body are separate, why and how do different mental states manifest themselves in different patterns of neural activity in the brain?  It also appears that some laws of physics have to be violated for the immaterial to affect the material; precisely where and at what level this breakdown of physical laws is occurring appears mysterious.

A fourth alternative is that consciousness is a fundamental feature or property of reality and not an emergent feature that can be explained away in terms of purely physical processes (This alternative actually encompasses a range of views, see Chalmers (2002b) for a systematic exposition). One possibility is that while physical reality is causally closed, there are laws or properties over and above the physical that relate to consciousness, a view advocated by David Chalmers in his book (Chalmers, 1996). A closely related position, expressed for example by neuroscientist
 Christoff Koch from Caltech, is that consciousness is inherent in the fabric of reality (Koch, 2012), that consciousness is woven into the very nature of the cosmos. How should we think about this possibility? While sometimes consciousness has been compared to a fundamental property of matter such as electrical charge, electrical charge can be associated with elementary particles and it is not clear that the same can be said about consciousness. Should we say that each particle carries some amount or quantum of consciousness? Since the brain is made up such a gigantic number of elementary particles, how then can we make sense of the cohesiveness and unitarity usually associated with human consciousness? There is also a second issue to consider. We believe that our subjective experiences and states can guide our action, but how can that be so if the physical universe is causally closed, and consciousness is an additional property of matter? This point has been discussed in detail by Chalmers (1996), but nevertheless leaves us uneasy.  Also there appears something disconcerting in the idea that even in principle we can never know or test whether a creature (or machine or robot) is conscious since presence or absence of consciousness will not influence the physical behavior of that system. Despite these apparent problems, this appears as the most promising alternative, and I propose that if we want to understand and make sense of this alternative more clearly, we should probe more deeply the nature of physical reality itself. 

      I will wrap up the discussion in this section with one comment. Why is it so difficult to find a self-consistent framework to elucidate the mind-brain relationship? The difficulty, it appears, is the self-contradictory nature of the relationship between the subjective and objective. On the one hand, as discussed before, a strong case has been made for an unbridgeable gap between the two. On the other hand, the subjective and objective also appear to interpenetrate strongly: as stated earlier, different mental and emotional states can be distinguished in terms of the patterns of neural activity. We know that drugs and diseases that affect the functioning of brain affect also our subjective states. Moreover, we believe that our subjective experiences and subjective mental states guide our actions, thus the subjective seems capable of acting on the objective. Finally (and paradoxically) we also have some expectation that the objective physical universe is causally closed. Can a philosophical framework reconcile these apparently contradictory aspects of the relationship? Perhaps surprisingly, we will answer in the affirmative.

\section{Physical reality and quantum mechanics}
 In everyday living, we assume the external world that we observe and experience corresponds to an objective external reality existing independent of us. However, a moment's reflection should convince us that what we view as objective external reality is in fact our internal mental representation of such a reality, based on incoming sensory information. Let me elaborate on this point by assuming here a conventional view of reality. At the time of writing, I perceive a table in front of me.  Presumably what is happening is that light reflected from the table is reaching my eyes and an image of the table is being created on the retina. This image generates electrical signals from photoreceptor cells, these signals travel to the brain and cause neuronal firings in the brain. Thus all that is reaching my brain are electrical signals that carry information from my sense organs. From these neuronal firings, somehow my brain is generating an image of the table. Thus, in some manner not quite understood, from their internal neurological states, our brains appear to create both our internal and our external world, with the external world being projected outside of the `self.' Thus what we are immediately aware of is not an objective external world but rather our internal mental representation of external reality. Logically, the most we can say is that this representation is self-consistent. It might be that external reality is organized exactly as we view it, but that is not {\it a-priori} a logical necessity. Nevertheless, it seems reasonable to ask whether we have any good reasons to believe that reality is not organized as it appears to us.

   Let us then ask what physics can tell us about the nature of reality. As we will see, quantum physics does fundamentally challenge our notion of reality (for popular accounts of quantum mechanics, and thought-provoking discussions, see for example, Herbert, 1985; Rosenblum et al., 2006; Scarani, 2006; Kumar, 2012). One of the most well-known aspects of quantum mechanics is the so-called wave-particle duality. As a result of dramatic developments in physics in the early twentieth century, physicists realized that light displays particle-like properties in addition to being electromagnetic waves, and moreover, that elementary particles such as electrons also display a wave nature. The wave nature of a quantum particle can be characterized by a function of space and time, known as the wavefunction, which represents the state of the particle and, for technical reasons, its deterministic time evolution as governed by Schr\"{o}dinger's equation is known as unitary evolution. Schr\"{o}dinger had speculated that an object's waviness was the smeared out object itself, so we could think of an object as a wavepacket: a burst of localized wave.  However, there was a problem with this interpretation since any observation trying to locate an electron at some spot either finds a whole electron or no electron at all, never part of an electron. Instead the square of the amplitude of the wave at any spot gives us the probability of finding the electron at that spot, if we look. Thus the accepted terminology is to call the wave a probability wave and to call the amplitude of the wave the probability amplitude. There is however a subtlety here concerning the difference between finding the particle at some spot versus the particle being there even if we have not performed an observation or measurement. To highlight why this difference is crucial and central to the mystery of quantum mechanics, let us consider a particular experimental set-up.

\begin{figure}
\begin{center}
\resizebox{10cm}{!} {\includegraphics{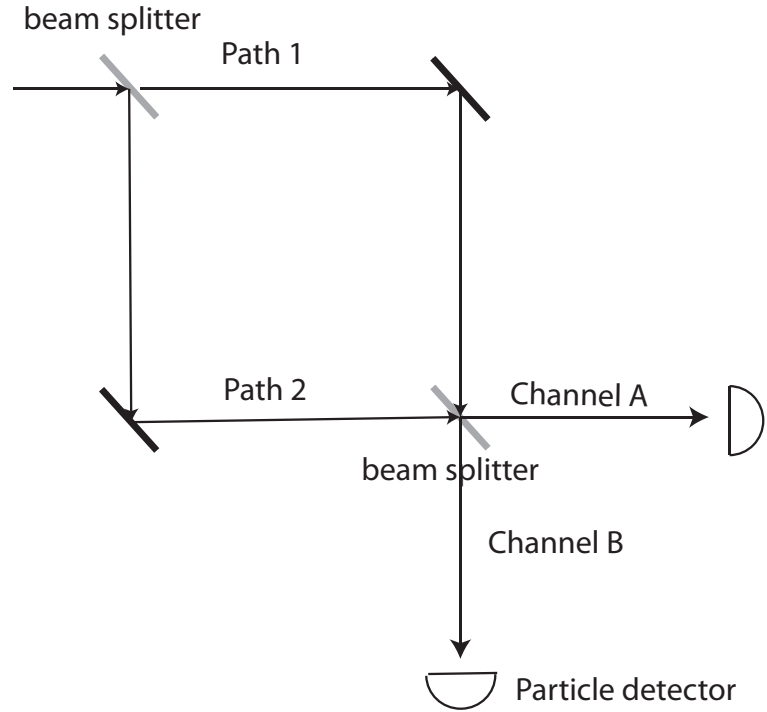}}
\caption{Experimental set-up for the Mach-Zehnder interferometer.  }
\label{Fig1}
\end{center}
\end{figure}

Consider a set-up where a beam of quantum particles, which could be photons, electrons, neutrons, or atoms, falling on a semi-transparent mirror where part of the beam gets reflected and part gets transmitted. This mirror will be called a beam-splitter, and can be so constructed that the intensity of transmitted beam equals the intensity of the reflected beam. Moreover the incident intensity is turned down so low that at any instant of time we can reasonably consider at most one particle in the equipment. For each particle falling on the mirror, there is a $50\%$ probability that it is transmitted and $50 \%$ that it is reflected. Two completely reflecting mirrors are set up along the paths (as shown in figure), directing the two beams towards a second semi-transparent mirror that also reflects $50\%$ and transmits $50\%$ of the particles. Following reflection or transmission at the second beam-splitter the particles could emerge along two possible channels/paths, represented in the figure as channels A and B. Particles along channel A are reflected at one semi-transparent mirror and transmitted through the other, whereas those along channel B are reflected by both or transmitted by both. Such an arrangement is called the Mach-Zehnder interferometer and highlights the weirdness of quantum phenomena. Classical probability theory predicts that $50\%$ of the particles come out along channel A and $50\%$ along channel B. This is because half the particles following path 1 get reflected to channel A and half transmitted to channel B, and similarly for particles following path 2. The observed results are dramatically different. All the particles come out along channel A, none along channel B (Granger et al., 1986); (for a popular and splendid discussion of this experiment, see Scarani, 2006). Quantum mechanics explains this by positing that the wavefunction of each individual particle hitting the beam-splitter splits in two, with part being reflected and part being transmitted, and predicts the correct result as an interference effect between the two parts of the wavefunction at the second beam splitter (the two parts of the wavefunction cancel out along channel B, leading to zero probability of particles being found in this channel). As a further check, quantum mechanics predicts that if the path-length of one of the paths is altered slightly (by placing additional mirrors along the path), the phase relation between the two parts of the wavefunction is altered and then particles will come out along both channels. The predictions of quantum mechanics are borne out quantitatively by experiments, providing a brilliant confirmation of quantum mechanics. The only way we can understand the experimental results is by allowing the particles to follow both paths (the alternative, that particles are guided by pilot waves that determine their trajectories, falls in the class of hidden variable theories that we will discuss briefly later in the paper). However, if we try to measure which path the particle followed by placing particle detectors, or in the case of atoms, by shining light on one of the paths, we will find the particle in its entirety always on path 1 or path 2, but the interference effect also disappears simultaneously and particles emerge along both channels.

      The upshot of the above argument is that the quantum particle simultaneously takes path 1 and path 2 unless we try to measure which path it is in. The so-called collapse of the wavefunction upon measurement is the centerpiece of the standard textbook approach to quantum mechanics developed by Niels Bohr and his group in the 1920s/30s, and is known as the Copenhagen interpretation (Wheeler and Zurek, 1983).  What constitutes a measurement? What is clear is that interaction between two quantum particles does not constitute a measurement since such an interaction is governed by Schr{\"o}dinger's equation, which is deterministic and does not lead to any collapse of wavefunction. The Copenhagen approach is to sharply demarcate between small things governed by quantum mechanics and Schr{\"o}dinger's equation, and large classical objects such as a measuring device. Schr\"{o}dinger's equation applies to quantum systems as long as we're not measuring. Measurement involves interaction with a classical object and the collapse of a wavefunction during measurement is {\emph{not}} governed by Schr{\"o}dinger's equation. Where does the boundary between a quantum and classical object lie?  This is a particularly pertinent question since continued development in technology allows physicists to experimentally study systems at all length-scales between the atomic and the macroscopic. And nowhere have we so far found direct evidence of Schr\"{o}dinger's equation breaking down. 

    So is there a way to figure out when a system behaves as a quantum system and when it behaves classically? In regards to superposition of states and quantum interference, the demarcation between the classical and the quantum arises, at least in part, from the phenomenon of decoherence. To understand decoherence, we will first touch upon the idea of entanglement, an idea that dates back to an important paper by Einstein, Podolsky, and Rosen (EPR; 1935). What they demonstrated is that quantum mechanics allows us to prepare two (or, in general, multiple) quantum particles such that their states are correlated, so that a measurement on one will directly effect the measurement on the second particle even though the two particles may be separated over an enormous distance. We say that the states of the two particles are entangled so that we cannot assign well-defined states to the individual particles, and entanglement has been demonstrated experimentally by numerous experiments. In this sense, quantum mechanism has an inherent holism (the idea of non-separability of quantum constituents) built into its mathematical structure.

       Given the inherently nonlocal mathematical structure of quantum mechanics, if the theory of quantum mechanics is complete and if we assume the position of realism outlined in the introduction, this appears to imply that reality in intrinsically nonlocal. Nonlocality, however, challenges our notion of cause and effect since from Einstein's theory of relativity we know that if disturbances or signals can propagate faster than the speed of light, they can also propagate backwards in time. 
 It appeared that a way out was provided by hidden variable theories, which posited that quantum description is incomplete, that is, that the wavefunction does not provide a complete description of the true state of the particle, and that there are hidden variables related to the state of a quantum particle which affect the outcome of a measurement. In a seminal article by John Bell (1982), he proved a theorem, now known as Bell's theorem, which makes it possible to experimentally differentiate between quantum mechanics and any local realist interpretation of quantum mechanics. The predictions of quantum mechanics have been convincingly borne out by experiments (Aspect et al., 1981) ruling out local hidden-variable theories. We are thus left with only the following alternatives: either to abandon realism or the notion of locality, or both (D'espagnat, 2003).  

     Let us return now to the quantum-classical gap. For an object to behave classically, the quantum state of the object entangles with that of the environment. The system can no longer be described by a pure quantum state. Provided Schr\"{o}dinger's equation does not break down, if the states of two particles get entangled their states could remain entangled for ever even though they may have stopped interacting directly with each other, indicating that it is reasonable to believe that the whole universe is in one entangled quantum state, with the quantum state of any object entangled with that of all others. So it seems like there are two levels of reality: the classical and the quantum; how do we relate them? The concept of decoherence (Zurek, 2003; Dass, 2003) is often invoked in this context. To understand the idea of decoherence, let us go back to the experiment with the beam splitter (Fig.~1). Consider a beam of atoms that is being split and atoms along one of the paths are allowed to interact with the environment; thus, that component of the wavefunction entangles with that of the environment. So the `phase' relationship between the two parts of the wavefunction (along paths 1 and 2) changes from instant to instant, implying in practical terms that the interference pattern will change from instant to instant. If we do an average over some time period, $Dt$, (or average over a few trials, as is necessary to obtain any pattern) the interference effect will disappear, and the system will appear exactly as if each particle were at random either in path 1 or path 2.  Nevertheless if we were earlier saying that the particle was actually in both paths, it appears unreasonable to now say that the particle is just going through one path independent of observation, thus not resolving the measurement problem. 
     
     To appreciate the difficulty, think of a classical random event such as a coin toss. While we may not know beforehand whether the outcome will be a head or tail, we expect it will have a definite outcome independent of whether anyone observes it. If we could imagine a quantum coin, then if `head' and `tail' are two allowed states, any superposition of the two states is an equally valid state. And unless the coin was specially placed in a head or tail state, we accord no special status to the head or the tail state and have no reason to believe the coin would be purely in the head state or purely in the tail state. Ignoring all the other microscopic degrees of freedom of the coin, physics tells us that in the classical limit the state of the coin entangles with that of the environment, and following the coin toss, this entangled state has two  relatively separate (decoherent) branches: one corresponding to a head and the other corresponding to a tail (Dass, 2003). So, in this sense, it appears that both the head and tail come up. If we take the concept of quantum states seriously, some prominent options are: (1) observation by a conscious observer collapses the entangled wavefunction to one of the two branches (Wigner, 1963), (2) the ``multiverse" splits into two branches/worlds, with a different outcome in each branch (many-worlds interpretation) (Everett, 1957) (3) there is an objective physical mechanism for wavefunction collapse (Ghirardi et al., 1986) and (4) quantum description of physical reality is incomplete, for example David Bohm's nonlocal hidden variable theory (Bohm, 1952). However none of these interpretations have found universal acceptance. There are of course other possibilities that could be compatible with realism, of particular relevance is one of the  
newest interpretations known as quantum Bayesianism which treats the wave function as subjective, encoding the observer's personalist
Bayesian probabilities for the outcomes of his measurements on the system, while acknowledging the existence of an objective reality (Fuchs, 2010). Nevertheless, this approach also leaves the precise relation between objective reality and empirical reality unresolved.

\section{Two levels of reality}

      In this section, I will develop a self-consistent view of quantum  reality and consciousness, and in the next section I will discuss where this view stands in relation to the multiple interpretations of quantum mechanics. For our subsequent discussion, I will use the standard term ``unitary quantum mechanics'' (the word unitary, in this context, is used in a technical, mathematical sense, and not in its general English-language usage) to denote the theoretical notion that the physical states of physically isolated systems at all lengthscales can be treated as quantum states and their temporal evolution is deterministic and governed by Schr\"{o}dinger's equation (or, more generally, its relativistic counterparts).  Unitary quantum mechanics essentially posits that Schr\"{o}dinger's equation (or its appropriate relativistic counterparts) applies at all length scales from the very small all the way up to the scale of the total universe. Given the remarkable success of quantum theory, and the fact that we have not yet found a single concrete context where we can show that unitary time development of the quantum state (as governed by Schr\"{o}dinger's equation) breaks down, unitary quantum mechanics seems to be a reasonable starting point. The case for unitary quantum mechanics is bolstered by examples of the macroscopic manifestations of quantum mechanics as in superconductivity and Bose Einstein condensation. One of the most dramatic confirmations of unitary quantum mechanics comes from an experimental study that was related to a 1985 paper entitled  `Quantum mechanics versus macroscopic realism: is the flux there when nobody looks?',  where A.J. Leggett and A. Garg (1985) suggested a quantitative test to determine whether a macroscopic object would at all times be in one of its distinct macroscopic states or whether quantum mechanics and quantum superposition would prevail. Twenty-five years later Agustin Palacios-Laloy and colleagues reported an experimental realization of the Leggett-Garg test (Palacios-laloy et al., 2010; Mooij, 2010) where using a superconducting circuit similar to the one proposed originally, they demonstrated that the behavior of their `macroscopic object' clearly follows the predictions of quantum mechanics. While unitary quantum mechanics is usually associated with Everettian many worlds interpretation (Everett, 1957),  in this section we will view unitary quantum mechanics with fresh eyes, and return to the many worlds interpretation in the following section. 

    What can we say about quantum reality referenced by the mathematical structure of unitary quantum mechanics? It turns out to be easier to describe quantum reality in terms of the negative rather then positive. What we know is that this underlying reality is marked by the absence of distinct, spatially localized objects as we experience at the classical level. More generally, we have no localized particles as well at this level, what we call quantum particles should be more legitimately thought of as excitations of non-local `fields' (Hobson, 2013). The most striking feature is the notion of non-separability, as discussed earlier. Quantum non-separability precludes in a novel way the possibility of defining individual objects independently of the conditions under which their behavior is manifested. It relates to the idea of entanglement in quantum mechanics: if the states of two particles are entangled, we cannot assign well-defined states, even in principle, to the individual particles. In this sense, reality is intrinsically non-local and interconnected at this level, though, technically, interactions are still local, thus ruling out the possibility of faster-than-light signaling.

       Here is what unitary quantum mechanics seems to imply: that there are two levels of reality, one corresponding to our experienced classical reality and the second, an underlying level of reality, referenced by the mathematical structure of quantum mechanics. Our classical world appears as one of the relatively independent branches of the underlying reality. If there existed only a single observer, saying that the observer is only aware of a part or a branch of total reality, while maybe philosophically unsatisfactory, poses no great logical challenge. The logical problem is to account for intersubjective agreement about external reality among multiple observers.  Let us, at this stage in the argument, remind ourselves of the discussion on consciousness. We were lead to the alternative that consciousness should be treated as being an irreducible component of nature, one possibility being that it is inherent in the fabric of reality. Since empirical reality is an emergent structure, applying fundamental rules or properties at this level does not seem to make sense and thus, if we were to treat consciousness as being fundamental, it seems entirely reasonable to assume that consciousness is an irreducible component of the underlying level of reality. Let us then posit some form of dualism at the underlying level, some component or principle in addition to physical reality governed by the laws of quantum mechanics, but which does not directly affect the unitarity of quantum mechanics. In this paper, I will use the phrase `underlying protoconsciousness' to refer to this irreducible component of underlying reality that is distinct from the physical aspects and provides the ground for phenomenal consciousness. 
       Bearing in mind that at this level, physical reality is fundamentally non-local, that there are no demarcated, localized objects with which we can associate consciousness, we can expect underlying protoconsciousness to be nonlocal as well. 

  We could then postulate that awareness of the particular quantum branch that corresponds to empirical reality fundamentally arises from this deeper underlying level of consciousness. We now see how the two pieces, the discussion of the nature of consciousness and that of the nature of reality, have to fit together. Our picture gives us both a framework for thinking about the relationship between mind and matter, and also provides us with an explanation of the role of the observer in quantum mechanics, in terms of an underlying level of reality. At the underlying level the temporal evolution of the total quantum state is deterministic and unitary, randomness and probability arise at the empirical reality and can be associated with the lack of knowledge of full reality, which fits in within a Bayesian framework of probability. At the underlying level, protoconsciousness should be treated as distinct from the physical aspect of reality and does not directly act upon the physical, thus avoiding the problems of Cartesian dualism. This leaves us with a view of reality, which, while to me entirely consistent with a scientific framework, does overlap with ancient religious and mystical traditions, both Eastern and Western. The proposed view overlaps with the views proposed by Goswami (1995), and could be characterized as a form of veiled realism propounded by d'Espagnat (2003, 2006): the notion of a true objective reality which is veiled from direct experience. It constitutes a middle way between materialism/physicalism and radical idealism. 

     How, within this view, do we treat the ontological status of empirical physical reality? An analogy with the ontological status of rainbows within a classical physical world helps clarify the issue, as highlighted by d'Espagnat (2006). Within classical physics, the question of how a rainbow forms poses no great conceptual challenge. Nevertheless the question of the ontological status can easily become confusing. Within a classical, mechanistic approach, a rainbow, obviously, may not be considered to be an object per-se, as its position (and even existence) is clearly dependent on the position of the observer. Should we say that the rainbow is non-existent and a creation of the observer? That clearly is not true either, rooted as the formation of the rainbow is in objective physical processes such as reflection of light. We could propose the existence of an infinite multitude of rainbows, where the observer picks out one from this infinite multitude depending on his or her location; but it is not clear whether such a view serves more to mystify a reasonably straightforward physical phenomenon. We face a similar problem when trying to clarify the ontological status of empirical physical reality. On the one hand, within our view it is rooted in objective observer-independent reality, existing as a branch of the underlying physical reality. On the other hand, it is also true that underlying consciousness is implicated in the process of selection of a particular branch as empirical reality. This, of course, should not be taken to mean that empirical physical reality 
depends on our individual consciousness; rather, I suggest interpreting this in terms of an intertwining of physical reality and consciousness at the empirical level.

   There is also an interesting issue here regarding wavefunctions. While we started with the assumption that quantum states are real, we found that the universe is in an entangled state, and since in unitary quantum mechanics there is no objective collapse of the quantum state, typically we cannot assign objective well-defined states to an individual particle even after a measurement. Thus the wavefunction represents {\it our knowledge} regarding the state of the particle, rather than the true state. The true universal wavefunction is hidden and not experimentally accessible, even in principle. In this sense, our view of unitary quantum mechanics could be characterized as a nonlocal hidden variable theory and indeed bears some resemblance to late physicist David Bohm's philosophical vision of reality (Bohm and Hiley, 1993).   

We now revisit the mind-body problem. In the proposed framework, both matter and mind at the experiential level emerge (not in any temporal sense) from the underlying reality -- a reality that is characterized by fundamental unity and non-separability of all of nature.  In that sense this position is not dualism in the conventional Cartesian sense, but neither is it monism in either the materialist or idealist sense (it could be regarded as a version of dual-aspect or neutral monism) since we postulate that the physical and the mental are two distinct aspects or attributes of underlying reality.  We could somewhat loosely regard underlying protoconsciousness as the link between physical cognitive processes of the brain at the empirical level and our individual conscious minds. While our framework does not preclude the possibility of neural correlates of consciousness (Tonomi and Koch, 2008) it does preclude the possibility of logically deducing facts about the quality of subjective experience directly from the neural correlates and physical principles. In regards to causal closure,
by insisting on unitary evolution of quantum states, we have ensured that the physical universe is causally closed at the underlying level but not at the empirical level. If the lack of causal closure appears disturbing, it is important to remember that the laws of physics that appear at the empirical level, such as Newton's laws of motion, should not be treated as fundamental laws, they are approximate and statistical laws that emerge from the underlying quantum dynamics. Thus while they hold, often with high precision, for a wide range of macro-phenomena, situations where, for example, microscopic fluctuations might get amplified to a macro-level would not be accurately or completely described by such laws. Finally, since consciousness participates in the generation of empirical reality while, simultaneously, physical neurological processes are connected to our conscious mental states, we find an interpenetration of the objective and subjective at the empirical level; we encounter this interpenetration both in our search for a resolution of the mind body problem and also for a satisfactory interpretation of quantum mechanics.  

 While my proposed view resembles strongly that of the philosopher Baruch Spinoza, let me here briefly relate my view to the philosophy of Immanuel Kant, a central figure in modern Western philosophy. My view of the two levels of reality parallels Kant's distinction between phenomenal (thing-as-perceived) and noumenal (thing-in-itself) reality (Kant, 1968), though my line of reasoning for arriving at this view is distinct from his. It is a nontrivial statement that the mathematical structure of quantum physics appears to penetrate, at least partially, into noumenal reality. This helps explains why, as physics attempts to probe more deeply into the nature of reality (as in high-energy physics, string theory etc.), physics also seems seems to move farther and farther away from our experience of reality.  
 For our discussion, however, what appears particularly relevant is Kant's notion of the ``transcendental unity of apperception,'' or, transcendental self, in short. Underlying our experiential empirical selves, Kant proposed the concept of a transcendental self that organizes and unifies all our experience in terms of an ``I'' that is experiencing them. It provides the condition for our conscious experience but, as Kant argues, is itself unknowable by introspection and cannot be described in empirical terms. I suggest that the notion of a transcendental self is intimately linked to the proposed notion of an
underlying protoconsciousness (in Kantian terms, the noumenal self), though I will leave the precise relation between the two as an open question at present.

       What I am offering in this paper is not a theory of consciousness, but rather the philosophical underpinnings for such a theory. A theory of consciousness would need to explain and elucidate why phenomenal consciousness is associated with certain processes and not with others; in our framework, it might also need to clarify further the relationship between our individual minds and underlying consciousness (just as quantum mechanics clarifies the relation between the underlying inter-connected physical reality and separable, localized, clearly demarcated, classical objects at the empirical level). Even without a detailed analysis, as would be required for such a theory, we can say something more about the relationship between mind and matter.  Our framework implies that reality manifests itself as a form of interactionist dualism (but not Cartesian dualism) at the empirical level, so that if a complex system or creature were to be conscious, its consciousness will be manifested in some form in its behavior; implying some behavioral signatures of consciousness. At the empirical level, the proposed framework implies that we treat the structure of conscious experience and physical neurological processes as two inter-dependent realms of research, but neither reducible to the other. It is important to emphasize that while this view does not, in my opinion, place consciousness outside the domain of science it does have implications regarding the aim of a scientific theory of consciousness. The aim of such a theory will not be to `explain' consciousness in physical or biological terms, but rather to provide an unified (and presumably causal) framework to integrate 
first-personal data (data about subjective experience, in the form of verbal reports) with third-personal data (data about physical neurological processes), as suggested by Chalmers (2010, Chapter 2). Finally, while for physicalism both the unity of apperception as well as the separation between the ``self" and the brain's representation of the external world remains puzzling, in our framework the unity of experience is a given; what still needs explanation is the notion of the self at the empirical level. This is a puzzle that, I believe, can  be addressed scientifically since an infant is likely not born with a sense of self but rather such a sense develops presumably from the infant's interaction with the physical and social environment.

\section{Addressing the reality crisis in quantum mechanics}
 
     In the words of Nick Herbert (1985) ``One of the best-kept secrets of science is that physicists have lost their grip on reality.'' This reality crisis manifests itself in the multiplicity of interpretations of quantum mechanics with no signs of reaching closure. Is accepting one or the other interpretation a matter of personal taste, or are there fundamental objective criteria for choosing between them? What is the relation between them and where does consciousness fit in within the multiplicity of interpretations?
        
      In order to address these questions, let us start with Everettian many world theory. The starting point of Everett's analysis was unitary quantum mechanics and the notion of an universal wave-function characterizing the physical state of the entire universe (Everett, 1957). This is not unlike our starting point in the previous section. In Everett's analysis, the focus then was to understand how reality would look to observers located within this universe. For objectivity, he replaced conscious observers by mechanical recording devices obeying natural laws, and studied the probabilistic correlation between such devices, thus addressing the question of inter-subjective agreement. He demonstrated how a deterministic process could appear random and irreversible to an observer. Nevertheless, despite the brilliance of his analysis, a fundamental problem remains. As discussed earlier, the universal wavefunction continues to branch, and each branch corresponds a classical/semi-classical world. From the viewpoint of quantum mechanics, all branches are equally real. Yet we experience a single empirical reality and we all agree about this reality. What decides which branch we experience as our empirical reality? Unitary quantum mechanics combined with physicalism, as assumed by Everett, can provide no answer to this question, thus ultimately unable to relate quantum reality back to our experience of reality (while there now exist several versions of many worlds interpretations, see Barrett (1999), Wallace (2012), to my knowledge none of them have successfully addressed the issue of a single empirical reality). It is due to this inability that the many-worlds interpretation seems particularly counterintuitive. As soon as we acknowledge the singularity of our experience and of empirical reality, it becomes clear that there has to be a selection at some level. It is this selection that is central to quantum mechanics. If unitary quantum mechanics holds and quantum mechanics provides a complete description of physical reality, then consciousness has to be involved in some way in this selection, since nothing in quantum mechanics confers any special status to one branch. 

    In light of the previous discussion, I propose the following statement as one of the central messages of quantum mechanics: ``our empirical reality is continuously being selected from a plurality of options at an underlying level of reality.'' While this is not an interpretation-free statement, it will help us organize the interpretations of quantum mechanics in terms of the two issues of `selection'  and  `underlying reality.' 
      
     Let us examine the issue of selection, which manifests itself in the interpretations in different ways, such as in the form of collapse of the wavefunction or as intrinsic probabilities.  A very general question we can ask is whether the selection can be understood in terms of a physical mechanism or do we need to invoke consciousness. If we accept the role of consciousness in the selection process, the issue of  inter-subjective agreement leads us rather directly to the notion of an underlying nonlocal protoconsciousness (one other possible alternative is an objective consciousness-induced collapse of the wavefunction which would then lead us to Cartesian dualism; I do not consider it explicitly here since I find it difficult to sustain as a self-consistent option; see Stapp, 2007,  for an alternative view). 
     Possible physical mechanisms for selection can be divided into two categories: deterministic or intrinsically random. If they are deterministic, as for example in Bohm's hidden variable theory, EPR paradox in conjunction with Bell's theorem implies the necessity of superluminal (faster-than-light) signaling, directly impacting our notion of cause and effect. Moreover, while such a theory might still be possible for non-relativistic quantum mechanics, extending it to relativistic quantum mechanics and to high energy physics remains a challenge. Bohm himself  considered his theory as an attempt to demonstrate that there was an alternative to the standard Copenhagen interpretation, which might lead to further clues, rather than as something final (Hiley, 2010). One other possibility raised by some researchers is that unknown processes at the extremely small lengthscales (Planck scales) would lead to a collapse of the wavefunction (Ghirardi et al., 1986). If these processes are deterministic, that however merely moves the issue of selection one step down the line without resolving it. Now let us turn to the possibility of intrinsic randomness. 
 There is a subtle argument against the notion of intrinsic randomness or intrinsic probability being rooted entirely in physical reality; the argument springs from considerations of the nature of time. In order to make sense of the notion of intrinsic randomness, we have to be able to root the temporal concept of the `present' entirely in physical reality; since intrinsic randomness would mean that even if the past and present are known with infinite accuracy, the future is not completely determined. However, the simultaneity of two spatially separated events has been shown to be frame-dependent by the theory of relativity, implying the impossibility of an objective global `present' or `now' (see also McTaggart (1908) for an independent argument against the objectivity of the present). If we cannot however root the concept of the `present' in objective reality, how can we make sense of objective intrinsic randomness?  To me, the only true alternative to a consciousness-based selection, as proposed here, might be a theory of objective collapse arising from some currently unknown physics presumably related to quantum gravity, as proposed, for example, by Penrose (1996).

          If we implicate consciousness in the selection process, it also becomes clear why realism is a choice. We could formulate quantum mechanics purely in terms of empirical reality, but the interpenetration of the objective and subjective makes it impossible for such a formulation to be entirely consistent with realism. The so-called collapse of the wavefunction after measurement, for example, is not some objective physical process but is related to our change in knowledge of the system. Thus how best to formulate quantum mechanics, at least in part, becomes a matter of choice and convenience. On the other hand we could formulate quantum mechanics in terms of an underlying reality, but have to bear in mind the role of consciousness in tying this underlying reality back to empirical reality. The question of what is the best representation of underlying reality: entangled states, or fields, or quantum information, is interesting but lies outside the purview of this paper. There is also a deep and fascinating question about what we {\emph{ can}} know about the state of underlying physical reality; recall that what we mean by wavefunction as it pertains to measurements relates to our knowledge of the state of the particle, and is not the true quantum state. Without undertaking a deep analysis here, we can nevertheless point to one aspect that we do have access to, namely, the number of distinct options available for each selection. In the technical language of quantum mechanics, this is the dimensionality of the Hilbert space,  and it is thus not mysterious that this quantity has been emphasized strongly in one of the newest interpretations known as quantum Bayesianism (Fuchs, 2010). Finally, our analysis could also have important implications for a third explanatory gap on the relation between physical and subjective time that I hope to explore in greater detail in the future. Consider, for example, the flow of time, a notion that is central to our experience of time but falls apart when we try to analyze it physically (see Davies, 1995, Chapter 12, for a discussion). Here again we find indirect evidence for the interpenetration of the subjective and objective. Our analysis suggests that the flow of time is related to the generation of empirical reality, and due to the role of underlying protoconsciousness in this process, it is not surprising that we cannot describe the flow of time in purely physical terms.

\section{Conclusions} 
  Phenomenal consciousness does not fit in naturally with our notion of physical reality. On the one hand, as briefly outlined in the paper, a strong case against physicalism has been developed by philosophers such as David Chalmers. On the other hand, finding a satisfactory alternative framework has also been challenging. In this paper, I have proposed an alternative framework, positing that underlying our experiential level of reality, there exists a deeper level of reality that is nonlocal and fundamentally interconnected. I have argued for a fundamental aspect of underlying reality that is distinct from the physical, which provides the basis for subjective conscious experience. In this framework, the physical world is causally closed at the underlying level, but not at the empirical level. Experienced empirical reality is emergent from this underlying reality and while we can for most practical purposes treat empirical physical reality as objective and observer-independent, the assumption of a consciousness-independent physical reality breaks down when we attempt to understand the relation between phenomenal consciousness and the physical brain processes.  I have demonstrated in this article how the proposed view of underlying reality could follow from quantum mechanics and how it resolves the interpretation problem in quantum mechanics. It also provides a resolution to the interlocking problems of mental causation and phenomenal consciousness (how phenomenal consciousness can act causally on physical neurological states) that were highlighted by Jaegwon Kim (2005). In the proposed view, subjective conscious states should be treated as being fundamental but intertwined with neurological states, and can act on the physical brain
   without violating any established physical laws or principles. The proposed view has important implications for a scientific theory of  consciousness. Such a theory presumably has to be rooted in empirical reality, just as quantum mechanics in its developmental phase had to be formulated in terms of a classical reality. At the empirical level, the relationship between subjective   
experience and physical brain processes should thus be treated as being just as fundamental as the elementary laws of physics, and the task of a theory of consciousness would be to provide a causal framework which would relate the two but without attempting to reduce one to the other. 
 
{\noindent {\bf Acknowlegements}} \\
I acknowledge stimulating discussions with and helpful suggestions from Prof. Les Blatt and Prof. Scott Hendricks. 

\bigskip

\centerline{\bf{REFERENCES}}

\bigskip

\small
\noindent
  Aspect, A.,  Grangier, P., and Gerard R. (1981) Experimental Tests of Realistic Local Theories via Bell's Theorem,  {\it Phys. Rev. Lett.}, {\bf 47}, pp 460-463.

\noindent
   Barrett, J.A. 1999. {\it The Quantum Mechanics of Minds and Worlds}, Oxford University Press.

\noindent
   Bell, J.S, (1982) On the impossible pilot wave, {\it Foundations of Physics}, {\bf 12}, pp 989-99. Reprinted in {\it Speakable and unspeakable in quantum mechanics: collected papers on quantum philosophy,} Cambridge University Press, 2004.

\noindent
   Bohm, D. (1952)  
A Suggested Interpretation of the Quantum Theory in Terms of ``Hidden" Variables. I.,  {\it Phys. Rev. }, {\bf 85}, pp 166-179. 

\noindent
Bohm, D. and Hiley, B. J. (1993), {\it The Undivided Universe: an Ontological Interpretation of Quantum Theory}, Routledge, London.

\noindent
Chalmers, D.J. (1995) Facing up to the problem of consciousness,
{\it Journal of Consciousness Studies}, {\bf 2}, pp  200-19.

\noindent
Chalmers, D.J. (1996), {\it The Conscious Mind: In Search of a Fundamental Theory}, New York: Oxford University Press. 

\noindent
Chalmers, D.J. (Editor). (2002a) {\it Philosophy of Mind: Classical and Contemporary Readings}, Oxford University Press.  
  
\noindent
Chalmers, D.J. (2002b) Consciousness and its Place in Nature, Published in {\it Philosophy of Mind: Classical and Contemporary Readings}. Oxford
University Press.

\noindent
Chalmers, D.J. (2010) {\it The Character of Consciousness (Philosophy of Mind)}. Oxford University Press. 

\noindent
  Dass, T. (2003) Measurements and Decoherence, preprint. URL: http://arxiv.org/abs/quant-ph/0505070.

\noindent
Davies, P.C.W.  (1995)  {\it About Time: Einstein's Unfinished Revolution}, Simon $\&$ Schuster.

\noindent
 D'Espagnat, B. (2003),
{\it Veiled Reality: An Analysis of Quantum Mechanical Concepts}, Westview Press, Boulder, Colorado. 

\noindent
 D'Espagnat, B. (2006) {\it On Physics and Philosophy}, Oxford University Press.

\noindent
  Einstein, A., Podolsky, B., Rosen, N. (1935) Can Quantum-Mechanical Description of Physical Reality Be Considered Complete?, {\it Physical Review}, {\bf 47}, pp 777.

\noindent
  Everett, H. (1957) Relative State Formulation of Quantum Mechanics, {\it Rev. Mod. Physics}, {\bf 29}, pp 454-462.
  
\noindent
Fine, A. (1996) {\it The Shaky game: Einstein, realism and the quantum theory}, 2nd ed. Chicago: University of Chicago Press.

\noindent
Fuchs, C.A. (2010) QBism, the Perimeter of Quantum Bayesianism. URL: 
http://arxiv.org/abs/1003.5209.

\noindent
Ghirardi, G.C., Rimini, A. and Weber T. (1986)   Unified dynamics for microscopic and macroscopic
systems, {\it Phys. Rev. D} {\bf 34}, pp 470-491.

\noindent
Goswami, A. (1995) {\it The Self-Aware Universe},  Tarcher.

\noindent
Grangier, P., Roger, G., and Aspect, A. (1986) Experimental evidence for a photon anticorrelation effect on a
beam splitter: A new light on single-photon interferences, {\it Europhys. Lett.}, {\bf 1}, pp 173-179.  

\noindent
 Herbert N. (1985) {\it Quantum Reality, beyond the new physics}, Anchor Press.

\noindent
Hiley, B.J. (2010) Some remarks on the evolution of Bohm's proposals for an alter-
native to standard quantum mechanics. preprint, [Online], http://www.bbk.ac.
uk/tpru/BasilHiley/History\_of\_Bohm\_s\_QT.pdf.

\noindent
Hobson, A. (2013)  There are no particles, there are only fields, {\it Am. J. Phys.}, {\bf 81}, pp 211-223.

\noindent
Jackson, F. (1982) Epiphenomenal qualia, {\it Philosophical Quarterly}, {\bf 32}, pp 127-136.

\noindent
Kant, I., (1968), {\it Kant's Critique of Pure Reason}, Norman Kemp Smith (trans.). New York: St. Martin's. 

\noindent
Kim, J. (2005) {\it Physicalism, or Something Near Enough}, Princeton, NJ: Princeton
University Press.

\noindent
Koch, C. (2012) {\it Consciousness: Confessions of a Romantic Reductionist}, The MIT Press.

\noindent
  Kumar, M. (2011) {\it Quantum: Einstein, Bohr, and the Great Debate about the Nature of Reality}, W. W. Norton $\&$ Company; Reprint edition.

\noindent
Laureys, S.,  and Tononi, G., (Editors)  (2008)
{\it The Neurology Of Consciousness: Cognitive Neuroscience and Neuropathology}, Academic Press. 

\noindent
Leggett, A.J., Garg, A. (1985) Quantum Mechanics versus macroscopic realism: is the flux there when nobody looks?, {\it Phys. Rev. Lett.}, {\bf 54}, 857-860.

\noindent
Linde, A. (2002)  Inflation, Quantum Cosmology and the Anthropic Principle, In
{\it Science and Ultimate Reality: From Quantum to Cosmos}, J. D. Barrow, P.C.W. Davies, $\&$ C.L. Harper eds., Cambridge University Press. 
URL: http://arxiv.org/abs/hep-th/0211048v2 .

\noindent
 McTaggart, J.E. (1908)
  The Unreality of Time, {\it Mind: A Quarterly Review of Psychology and Philosophy}, {\bf 17}, pp 456-473.

\noindent
Mooij, J.E. (2010)  No moon there, {\it Nature Physics} {\bf 6}, pp 401-402. 

\noindent
Nagel, T. (1974) What Is It Like to Be a Bat?, {\it The Philosophical Review}, {\bf 83}, pp 435-450.

\noindent
Palacios-laloy, A. {\it et. al}. (2010) 
`Experimental violation of a Bell's inequality in time with weak measurement, {\it Nature Physics}, {\bf 6}, pp 442-447.

\noindent
Penrose, R. (1996)  On gravity's role in quantum state reduction, {\it Gen. Rel. Grav.} {\bf 28}, pp 581-600.

\noindent
 Ramachandran, V.S. and Hubbard, E.M. (2003)
The Phenomenology of Synaesthesia, {\it Journal of Consciousness Studies}, {\bf 10}, pp 49-57.

\noindent
 Rosenblum, B. and Kuttner, F. (2006) {\it Quantum Enigma: Physics Encounters Consciousness}, Oxford University Press, USA; Reprint edition.

\noindent
Scarani, V. (2006) {\it Quantum Physics - A First Encounter}, Oxford University Press. 

\noindent
Stapp H.P.  (2007) {\it Mindful Universe: Quantum Mechanics and the Participating Observer}, Springer.

\noindent
  Tononi, G. and Koch, C. (2008) The Neural Correlates of Consciousness: An Update, {\it N.Y. Acad. Sci.} {\bf 1124}, pp 239-261.

\noindent
Wallace, D. 2012.
{\it The Emergent Multiverse: Quantum Theory according to the Everett Interpretation}, Oxford University Press.  

\noindent
  Wheeler, J.A.,  and  Zurek, W.H.,  (eds). (1983) {\it Quantum Theory and Measurement}, Princeton University Press.

\noindent
 Zurek, W.H. (2003) Decoherence, einselection, and the quantum origins of the classical, {\it Rev. Mod. Phys.} {\bf 75}, pp 715.

\end{document}